\documentclass[12pt]{article}

\usepackage{amstext,amsmath,amssymb,amsfonts}
\usepackage{cite}
\usepackage{graphicx}
\usepackage{epsfig}

\usepackage[all]{xy}
\usepackage{tabularx}

\usepackage{latexsym}
\newcommand{\be}{\begin{equation}}
\newcommand{\en}{\end{equation}}

\newtheorem{thm}{Theorem}
\newtheorem{cor}[thm]{Corollary}

\newtheorem{defi}{Definition}[section]
\newtheorem{lem}[defi]{Lemma}

\newcommand{\bedefin}{\begin{defi}}
\newcommand{\findefi}{\end{defi} \medskip}

\newcommand{\betheo}{\begin{theorem}$\!\!${\bf \,\,\,}}
\newcommand{\entheo}{\end{theorem}}
\newcommand{\enth}{\end{theorem}}

\newcommand{\becor}{\begin{cor}$\!\!${\bf .}}
\newcommand{\encor}{\end{cor}}

\newcommand{\belem}{\begin{lem}$\!\!${\bf }}
\newcommand{\enlem}{\end{lem}}

\newcommand{\bea}{\begin{eqnarray}}
\newcommand{\ena}{\end{eqnarray}}

\newcommand{\ee}{\end{equation}}

\newcommand{\beano}{\begin{eqnarray*}}
\newcommand{\enano}{\end{eqnarray*}}

\newcommand{\bee}{\begin{enumerate}}
\newcommand{\ene}{\end{enumerate}}

\newcommand{\bei}{\begin{itemize}}
\newcommand{\eni}{\end{itemize}}

\newcommand{\betab}{\begin{tabular}}
\newcommand{\entab}{\end{tabular}}

\newcommand{\bd}{\begin{displaymath}}

\catcode `\@=11 \@addtoreset{equation}{section}

\catcode `\@=12

\title{\textbf{Conformal Symmetry Transformations\\
and Nonlinear Maxwell Equations}}

\author{\textit{Gerald A. Goldin $ {\!}^{\sf a}$,
Vladimir M. Shtelen $ {\!}^{\sf b}$, Steven Duplij $ {\!}^{\sf c}$}\\[10pt]
$^{\sf a}${\small Departments of Mathematics and Physics}\\
{\small Rutgers University, New Brunswick, NJ USA}\\
{\small \texttt{geraldgoldin@dimacs.rutgers.edu}}\\
$^{\sf b}${\small Department of Mathematics}\\
{\small Rutgers University, New Brunswick, NJ USA}\\
{\small \texttt{shtelen@math.rutgers.edu}}\\
$^{\sf c}${\small University of M\"unster, Germany}\\
{\small \texttt{duplijs@math.uni-muenster.de}}
}
\date{April 2, 2017}
\begin{document}

\maketitle
\thispagestyle{empty}

\setcounter{section}{0}
\setcounter{equation}{0}
\setcounter{figure}{0}
\setcounter{table}{0}
\setcounter{footnote}{0}



\begin{center}
{\it It is a pleasure to dedicate this article to M. Norbert Hounkonnou\\ on the occasion of his 60th birthday.}

\end{center}

\vspace{15pt}

\begin{abstract}
We make use of the conformal compactification of Minkowski spacetime $M^{\#}$ to explore a way of describing general, nonlinear Maxwell fields with conformal symmetry. We distinguish the inverse Minkowski spacetime $[M^{\#}]^{-1}$ obtained via conformal inversion, so as to discuss a doubled compactified spacetime on which Maxwell fields may be defined. Identifying $M^{\#}$ with the projective light cone in $(4+2)$-dimensional spacetime, we write two independent conformal-invariant functionals of the $6$-dimensional Maxwellian field strength tensors -- one bilinear, the other trilinear in the field strengths -- which are to enter general nonlinear constitutive equations. We also make some remarks regarding the dimensional reduction procedure as we consider its generalization from linear to general nonlinear theories.
\end{abstract}

\newpage

\begin{small}
\tableofcontents
\end{small}


\section*{Introduction}

It is well-known that in $(3+1)$-dimensional spacetime (Minkowski space), denoted $M^{(4)}$,
Maxwell's equations respect not only Poincar\'e symmetry, but conformal symmetry. But the
physical meaning of this conformal symmetry is still not entirely clear. A historical review is provided by Kastrup \cite{Kas2008}.

In our ongoing work, we have been investigating the characterization of general, nonlinear conformal-invariant Maxwell theories \cite{DupGolSht2014}. Our strategy is to make use of the identification of the conformal compactification $M^\#$ of Minkowski space with the projective light cone in $(4+2)$-dimensional spacetime $Y^{(6)}$ \cite{Dir1935}. Poincar\'e transformations, dilations and special conformal transformations act by rotations and boosts in $Y^{(6)}$. Nikolov and Petrov \cite{NikPet2003} consider a linear Maxwell theory in $Y^{(6)}$, and carry out a ray reduction and dimensional reduction procedure to obtain conformal-invariant theories in $M^{(4)}$. The result is a description of some additional fields that might survive in $M^{(4)}$. To handle nonlinear Maxwell theories, we allow the constitutive equations to depend explicitly on conformal-invariant functionals of the field strength tensors (with the goal of carrying out a similar dimensional reduction). This parallels, in a certain way, the approach taken by two of us in earlier articles describing general (Lagrangian and non-Lagrangian) nonlinear Maxwell and Yang-Mills theories with Lorentz symmetry in $M^{(4)}$ \cite{GolSht2001, GolSht2004}.

This contribution surveys of some of the key ideas underlying our investigation. A major tool is to focus on the behavior of the fields and the coordinates under conformal inversion. We introduce here the resulting ''inverse Minkowski space'' obtained via conformal inversion, and consider the possibility of defining Maxwellian fields independently on the inverse space. We also write  two independent conformal-invariant functionals of the Maxwell field strength tensors in $Y^{(6)}$ -- one bilinear, the other trilinear in the field strengths. These are the functionals which are to enter general nonlinear constitutive equations in the $(4+2)$-dimensional theory. We also make some remarks regarding the dimensional reduction procedure from six to four dimensions, as we consider its generalization from linear to general nonlinear theories.

\section{Conformal transformations and compactification}

\subsection{Conformal transformations in Minkowski space}

The full conformal group for $(3+1)$-dimensional Minkowski spacetime $M^{(4)}$, as usually defined, includes the following transformations. For $x = (x^\mu) \in M^{(4)}$, $\mu = 0, 1, 2, 3$, we have:

\noindent
translations:
\be
x^{\prime \, \mu}  = (T_b \,x)^\mu =  x^{\,\mu} - b^{\,\mu}\,;
\en

\noindent
spatial rotations and Lorentz boosts, for example:
\be
x^{\prime \, 0} = \gamma (x^0 - \beta x^1)\,, \,\,\, x^{\prime \, 1} = \gamma (x^1 - \beta x^0)\,, \,\,\,
 -1 < \beta = \frac{v}{c} < 1\,, \,\,\, \gamma = (1 - \beta^2)^{-\frac{1}{2}}\,;
\en

\noindent
or more generally,
\be
x^{\prime\, \mu} = (\Lambda x)^\mu = \Lambda_\nu^\mu\,x^\nu\, \quad \mathrm{(Einstein \,\, summation \,\, convention)}\,;
\en

\noindent
and dilations:
\be
x^{\prime \, \mu}  =  (D_\lambda \,x)^\mu = \lambda x^{\,\mu}\,, \quad  \lambda > 0\,;
\en

\noindent
all of which are {\it causal} in $M^{(4)}$. Let us consider {\it conformal inversion} $R$, which acts singularly on $M^{(4)}$, and breaks causality:
\be
x^{\prime \, \mu}  = (R\, x)^\mu =  x^{\,\mu}  / x_\nu x^{\nu}, \quad \mathrm{where}
\en
\be
x_\nu x^{\,\nu} =  g_{\mu \nu}  x_\mu  x^{\,\nu}\,, \quad
g_{\mu \nu}   =
\mathrm{diag}\,[1, -1, -1, -1]\,.
\en

Evidently $R^2 = I$. That is, neglecting singular points, conformal inversion is like a reflection operator: inverting twice yields the identity operation. Conformal inversion preserves the set of light-like submanifolds (the ``light rays''), but {not}
the causal structure. Locally, we have:
\be
g_{\mu \nu} dx^{\prime \, \mu}dx^{\prime \, \nu} = \frac{1}{(x_\sigma x^\sigma)^2} \,g_{\mu \nu} dx^{\mu}dx^{ \nu}\,.
\en

Combining inversion with translations, and inverting again, gives us the {\it special conformal transformations} $C_b$, which act as follows:
\be
x^{\prime \, \mu}   = (C_b \,x)^{\,\mu} =  (R T_b\, R\, x)^{\,\mu} =  (x^{\,\mu} - b^{\,\mu} x_\nu x^{\,\nu}) / (1 - 2b_\nu  x^{\,\nu} + b_\nu b^{\,\nu} x_\sigma x^{\,\sigma} )\,.
\en

\noindent
The operators $C_b$ belong to the conformal group, and can be continuously connected to the identity.

\subsection{Conformal compactification}

We can describe Minkowski space $M^{(4)}$ using {light cone coordinates}.
Choose a particular (spatial) direction in $\mathbf{R}^3$. Such a direction
is specified by a unit vector $\hat{u}$, labeled  (for example) by an appropriate choice of angles in spherical coordinates. A vector $\mathbf{x} \in \mathbf{R}^3$ is then labeled by angles and by the coordinate $u$, with $-\infty < u < \infty$, and $\mathbf{x}\cdot \mathbf{x} = u^2$.
With respect to this direction, we introduce the usual
light cone coordinates,
\be
u^{\pm} = \frac{1}{\sqrt{2}} (x^0 \pm u)\,.
\en

\noindent
Then for $x = (x^0, \mathbf{x})$, we have $\,x_\mu x^\mu \,=\,2u^+ u^-$; so under conformal inversion
(with obvious notation),
\be
u^{\,\prime \,+} \,=\, \frac{1}{2u^-}\,, \quad u^{\,\prime \,-} \,=\, \frac{1}{2u^+}\,\,.
\en

To obtain the conformal compactification $M^{\#}$ of the $(3+1)$-dimensional Minkowski space $M^{(4)}$, we formally adjoin to it the set $\mathcal{J}$ of
necessary ``points at infinity.'' These are the images under inversion of the light cone $L^{(4)} \subset M^{(4)}$ (defined by either $u^+ = 0$ or $u^- = 0$), together with the formal limit points of $L^{(4)}$ itself at infinity
(which form an invariant submanifold of $\mathcal{J}$).
Here $\mathcal{J}$ is the
well-known ``extended light cone at infinity.''
The resulting space
$M^{\#} \,=\,M^{(4)} \cup \mathcal{J}$ has the topology
of $S^3 \times S^1/Z_2$.

In the above, we understand the operators $T_a$, $\Lambda^\mu_\nu$, $D_\lambda$, $R$, and  $C_b$ as transformations of $M^{\#}$. Including these operators but leaving out $R$, we have what is often referred to as the ``conformal group,'' all of whose elements are continuously connected to the identity.
There are many different ways to coordinatize $M^{\#}$
and to visualize its structure, which we shall not discuss here.

\section{Inverse Minkowski space}

\subsection{Motivation and definition}

In the preceding construction, which is quite standard, there is a small problem with the units. We glossed over (as do nearly all authors) the fact that $x^{\,\mu}$ has the dimension of {\it length}, while the expression for $(R\,x)^{\,\mu}$ has the dimension of {\it inverse length}. Thus we cannot actually consider $R$ as a transformation on Minkowski space (or on compactified Minkowski space) without arbitrarily fixing a unit of length!

Furthermore, regarding the formula for $(C_b\,x)^\mu$, it is clear that $b$ must have the dimension of inverse length; but in the expression for $(T_b\,x)^{\,\mu}$, it has the dimension of length.

Kastrup (1966) suggested introducing a Lorentz-invariant ``standard of length'' $\kappa$ at every point, having the dimension of inverse length, and working with the dimensionless coordinates $\eta^\mu = \kappa x^\mu$ together with $\kappa$. This leads into a discussion of geometrical gauge properties of Minkowski space.

Let us consider instead the idea of introducing a separate ``inverse Minkowski space'' $[M^{(4)}]^{-1}$, whose points $z$ have dimension of inverse length. Then we can let $z^{\,\mu} = (\hat R \,x)^\mu = x^\mu / x_\nu x^\nu$ belong to $[M^{(4)}]^{-1}$. As before, in order to define $\hat R$ on the light cone, we shall need to compactify: first, to compactify $[M^{(4)}]^{-1}$ so as to include the image points of $\hat R$ acting on the light cone in $M^{(4)}$, and then to  compactify $[M^{(4)}]$, obtaining $[M^{\#}]^{-1}$ and $M^{\#}$. The two spaces are, of course, topologically and geometrically the same, with $\hat R : M^{\#} \to [M^{\#}]^{-1}$, and its inverse $\hat R^{-1}: [M^{\#}]^{-1} \to M^{\#}$, given by the same formula: $x^\mu = z^\mu / z_\nu z^\nu$.

Just as we have $T_a$, $\Lambda^{\mu}_{\nu}$, $D_\lambda$, and $C_b$ acting in $M^{\#}$ (allowing $a$ to have the dimension of length, and $b$ to have the dimension of inverse length), we now define corresponding transformations, $\tilde T_b$, $\tilde \Lambda^{\mu}_{\nu}$, $\tilde D_\lambda$, and $\tilde C_a$ acting in  $[M^{\#}]^{-1}$, using the same formulas as before, but with $z$ replacing $x$. Thus, $(\tilde T_b\,z)^{\mu} = z^{\,\mu} - b^{\,\mu}$, and so forth. Now,
\be
C_b = \hat R^{-1} \tilde T_b \hat R, \quad D_\lambda = \hat R^{-1} \tilde D_{1/\lambda} \hat R, \quad \Lambda = \hat R^{-1} \tilde \Lambda \hat R, \quad T_a = \hat R^{-1} \tilde C_b \hat R \,.
\en

\subsection{Conformal Lie algebra}

The well-known Lie algebra of the conformal group has $15$ generators, as follows:
\be
[P_\mu, P_\nu] = 0, \quad [K_\mu, K_\nu] = 0, \quad [P_\mu, d] = P_\mu, \quad [K_\mu, \textrm{d}] = -K_\mu, \nonumber
\en
\be
[P_\mu, J_{\alpha \beta}] = g_{\mu \alpha} P_{\beta} - g_{\mu \beta} P_{\alpha}, \,\quad\, [K_\mu, J_{\alpha \beta}] = g_{\mu \alpha} K_{\beta} - g_{\mu \beta} K_{\alpha}, \quad
\en
\be
[J_{\mu \nu}, J_{\alpha \beta} ] = \mathrm{(usual \,\, Lorentz \,\, algebra)}, \quad
[P_\mu, K_\nu] = 2(g_{\mu \nu} \textrm{d} - J_{\mu \nu}), \nonumber
\en

\noindent
where the $P_\mu$ generate translations, the $K_\mu$ generate special conformal transformations, the $J_{\alpha \beta}$ generate Lorentz rotations and boosts, and $\textrm{d}$ generates dilations.

Evidently the exchange $P_\mu \to K_\mu$, $K_\mu \to P_\mu$, $\textrm{d} \to -\textrm{d}$ leaves the Lie algebra invariant. This fact is now easily understood, if we think if it as conjugating the operators in $M$ with the operator $\hat R$ to obtain the generators of transformations in $[M^{\#}]^{-1}$:
\be
\tilde P_\mu = \hat R K_\mu \hat R^{-1},\quad  \tilde K_\mu = \hat R P_\mu \hat R^{-1}, \quad \tilde{\textrm{d}} = \hat R (-\textrm{d}) \hat R^{-1}, \quad \tilde J = \hat R J \hat R^{-1}.
\en

\subsection{Some comments}

To relate the original conformal inversion $R$ to $\hat R$, we may introduce an arbitrary constant $A > 0$, having the dimension of area. Let ${\hat A}: [M^{(4)}]^{-1} \to M^{(4)}$ be the operator  $x^{\,\mu} = A z^{\,\mu}$. Then define $x^{\prime \, \mu}  = (R_A x)^\mu = (\hat A \hat R x)^{\,\mu} = A x^{\,\mu} / x_\nu x^{\,\nu}$, for $A > 0$. Note that $R_A^{\,2} = I$, independent of the value of $A$.

Now (letting $b$ have units of length), we have $(R_A T_b R_A x)^{\mu} = (C_{b/A} x)^\mu$, and we can work consistently in the original Minkowski space and its compactification. The introduction of the constant $A$ parallels Kastrup's introduction of the length parameter $\kappa$.

However, it is also interesting not to follow this path, but to consider the {\it doubled, compactified Minkowski space} $M^{\#} \cup [M^{\#}]^{-1}$; i.e., the disjoint union of $M{^\#}$ and its inverse space. It is possible to define Maxwell fields on the doubled space, making use of the conformal inversion.

Finally we remark that a similar construction of an ``inverse spacetime'' can be carried out for the Schr\"odinger group introduced by Niederer (1972). The Schr\"odinger group consists of the Galilei group, dilation of space and time given by $D_{\lambda} (t, \mathbf{x}) = (\lambda^2 t, \lambda \mathbf{x})$, and additional transformations that can be considered as analogues of special conformal transformations. The latter transformations can be obtain as the result of an inversion, followed by time translation, and then inversion again. Here the inversion is defined by  $R: (t, \mathbf{x}) \to (-1/t, \mathbf{x}/t)$, with $R^{\,2}: (t, \mathbf{x}) \to (t, -\mathbf{x})$. Note that for the Schr\"odinger group, there is only a one-parameter family of transformations obtained this way, in contrast to the four-parameter family of special conformal transformations; the Schr\"odinger group is only $12$-dimensional, while the conformal group is $15$-dimensional.

Under inversion, the dimensions again change. Here, they change from time and space to inverse time and velocity respectively. Again one compactifies, and again we have the option to introduce a ``doubled spacetime," where now it is a compactified Galilean spacetime which has been doubled.

\section{Nonlinear electrodynamics: general approach}

\subsection{Motivation and framework for nonlinear Maxwell fields}
Let us write Maxwell's equations as usual (in SI units), in terms of the four fields $\mathbf{E}, \mathbf{B}, \mathbf{D}$, and $\mathbf{H}$:
\be
\nabla \times \mathbf{E} = - \frac{\partial \mathbf{B}}{\partial t}\,,\quad \nabla \cdot \mathbf{B} = 0\,,\quad
 \nabla \times \mathbf{H} =  \frac{\partial \mathbf{D}}{\partial t} + \mathbf{j}\,, \quad \nabla \cdot \mathbf{D} = \rho\,.
\en

\noindent
The {\it constitutive equations}, relating the pair $(\mathbf{E}, \mathbf{B})$ to the pair $(\mathbf{D}, \mathbf{H})$,
may be linear or nonlinear. Our strategy is to introduce general constitutive equations respecting the desired symmetry
at the ``last possible moment.''

Now the general nonlinear theory with Lorentz symmetry has constitutive equations of the form
\be
\mathbf{D} = M\mathbf{B} + \frac{1}{c^2}N\mathbf{E}\,, \quad  \mathbf{H} = N\mathbf{B} - M\mathbf{E}\,,
\en

\noindent
where $M$ and $N$ may depend on the field strengths via the two Lorentz invariants
\be
I_1 = \mathbf{B}^2 - \frac{1}{c^2}\mathbf{E}^2\,, \quad I_2 = \mathbf{B} \cdot \mathbf{E}\,.
\en

\noindent
In the linear case, $M$ and $N$ are constants.

Our initial motivation for studying nonlinear Maxwell theories with symmetry was to explore the existence of a Galilean limit \cite{GolSht2001}. It is well known that
taking a Galilean limit $c \to \infty$ in the linear case requires losing one of the time-derivative terms in Maxwell's equations, as described carefully by Le Bellac and L\'evy-Leblond \cite{LebLeb1973}. But in the general nonlinear case (allowing non-Lagrangian as well as Lagrangian theories), we showed that
all four Maxwell equations can survive intact. Here $I_1$ and $I_2$ survive, and can yield nontrivial theories in the $c \to \infty$ limit.

We remark here that introducing conformal symmetry in this context further restricts the invariants, leaving only the ratio $I_2 / I_1$ as an invariant.

In covariant form, Maxwell's equations are written (in familiar notation):
\be
\partial_\alpha \tilde{F}^{\alpha \beta} = 0\,, \quad \partial_\alpha G^{\alpha \beta} = \mathbf{j}^{\,\beta}\,,
\en

\noindent
where
\be
\tilde{F}^{\alpha \beta} = \frac{1}{2}\epsilon^{\alpha \beta \mu \nu}F_{\mu \nu} \quad \mathrm{and} \quad
 F_{\mu \nu} = \partial_\mu A_\nu - \partial_\nu A_\mu\,.
\en

\noindent
Here the constitutive equations relate $G$ to $F$ and $\tilde{F}$. With Lorentz symmetry, they take the general form
\be
G^{\mu \nu} \,=\, NF^{\mu \nu} + cM\tilde{F}^{\mu \nu} \,\equiv\, M_1 \frac{\partial I_1}{\partial F_{\mu \nu}} + M_2 \frac{\partial I_2}{\partial F_{\mu \nu}}\,,
\en

\noindent
where $M$ and $N$ (or, equivalently, $M_1$ and $M_2$) are functions of the Lorentz invariants $I_1$ and $I_2$:
\be
I_1 =  \frac{1}{2} F_{\mu \nu}F^{\mu \nu} \,, \quad I_2 = -\frac{c}{4}F_{\mu \nu}\tilde{F}^{\mu \nu}\,.
\en

\subsection{Transformations under conformal inversion}
Under conformal inversion, we have the following symmetry transformations of the electromagnetic potential, and of spacetime derivatives:
\be
A_\mu^{\prime}(x^\prime) = x^2 A_\mu(x) - 2x_\mu (x^\alpha A_\alpha)\,
\en
\be
\partial^\prime_\mu := \frac{\partial}{\partial x^\prime} = x^2 \partial_\mu - 2 x_\mu (x \cdot \partial)\,
\en

\noindent
where we here abbreviate $x^2 = x_\mu x^\mu$ and $(x \cdot \partial) = x^\alpha \partial_\alpha$. Then
with $F_{\mu \nu} = \partial_\mu A_\nu - \partial_\nu A_\mu$, we have:
\be
F^{\prime}_{\mu\nu}(x^\prime) = (x^2)^2 F_{\mu\nu}(x) - 2 x^2 x^\alpha (x_\mu F_{\mu\nu} + x_\nu F_{\mu\alpha})
\en

\noindent
and
\be
\Box^\prime = (x^2)^2 \Box - 4 x^2 (x \cdot \partial)\,,
\en

\noindent
where the d'Alembertian $\Box = \partial_\mu \partial^\mu$. Additionally, the $4$-current $j_\mu$ transforms by
\be \label{inversionj}
{j^{\,\prime}}_\mu (x^\prime) = (x^2)^3 j_\mu (x)\,-2(x^2)^2 x_\mu (x^\alpha j_\alpha(x))\,.
\en

These transformations define a symmetry of the (linear) Maxwell equations,
\be \label{Maxwell}
\Box A_\nu - \partial_\nu (\partial^\alpha A_\alpha) =  j_\nu\,.
\en

\noindent
That is, if $A (x)$ and $j (x)$ satisfy (\ref{Maxwell}), then $A^\prime (x^\prime)$ and $j^{\prime} (x^\prime)$ satisfy the same equation with $\Box^{\,\prime}$ and $\partial^\prime$ in place of $\Box$ and $\partial$ respectively. Combining this symmetry with that of the Poincar\'e transformations and dilations, we have the symmetry with respect to the usual conformal group.

But note that the symmetry under conformal inversion can be interpreted to suggest not only a relation among solutions to Maxwell's equations in $M^{\#}$, but also the definition of {new} Maxwell fields on the inverse compactified Minkowski space $[M^{\#}]^{-1}$.

\subsection{Steps toward general nonlinear conformal-invariant electrodynamics}

We see the remaining steps in constructing general, nonlinear conformal invariant Maxwell theories (both Lagrangian and non-Lagrangian) as the following. Identifying $M^{\#}$ with the projective light cone in the $(4+2)$-dimensional space $Y^{(6)}$, we write Maxwell fields in $Y^{(6)}$, and constitutive equations in $Y^{(6)}$. The constitutive equations depend only on conformal-invariant functionals of the Maxwell fields in $Y^{(6)}$, which we identify. To restrict the theory to the projective light cone, we then carry out a dimensional reduction procedure, as discussed by Nikolov and Petrov \cite{NikPet2003}. In doing this we make use of the ``hexaspherical space'' $Q^{(6)}$ -- transforming all the expressions to hexaspherical coordinates, and proceeding from there.

\section{Related (4+2)-dimensional spaces}

In this section we review the $(4+2)$-dimensional spaces $Y^{(6)}$ and $Q^{(6)}$, highlighting how conformal inversion acts in these spaces.

\subsection{The space $Y^{(6 )}$}

For  $y \in  \mathbf{R}^6$, write $y = (y^m), m = 0, 1, \dots,5$,
define the
flat metric tensor $\eta_{mn} = \mathrm{diag} [1, - 1, - 1, - 1; - 1, 1] $,
so that
\be
y_m y^m  \,=\, \eta_{mn} y^m y^n \, =\, (y^0 )^2 - (y^1)^2 - (y^2 )^2 - (y^3 )^2
                     - (y^4)^2 + (y^5)^2\,.
\en

This is the space we call $Y^{(6 )}$.  The light cone $L^{(6)}$ is then the submanifold specified by the condition,

\be
y_m y^m  = 0\,,\,\,\, \mathrm{or} \,\,\,
    (y^1)^2 + (y^2)^2 + (y^3)^2 + (y^4)^2  \,=\, (y^0)^2 + (y^5)^2\,.\label{lc6}
\en

\noindent
To define the projective space $PY^{(6)}$ and the projective light cone $PL^{(6)}$,
consider  $y  =  (y^m ) \in  Y^{(6)}$, and define the projective equivalence relation,
\be
 (y^m ) \sim (\lambda y^m ) \,\,\,  \mathrm{for} \,\,\, \lambda \in \mathbf{R}, \,\lambda \neq 0\,.
 \en

\noindent
The equivalence classes $[y]$ are just the {\it rays} in $Y^{(6)}$; and $PY^{(6)}$ is this space of rays.

\noindent
To describe the projective light cone $PL^{(6)}$, we may choose one point in each ray in $L^{(6)}$.
Referring back to Eq. (\ref{lc6}), if we consider
\be
(y^1)^2 + (y^2)^2 + (y^3)^2 + (y^4)^2  \,=\, (y^0)^2 + (y^5)^2  \,=\, 1,
\en

\noindent
we see that we have $S^3 \times S^1$. But evidently the above condition selects two points in each ray; so $PY^{(6)}$
can in this way be identified with (and has the topology of)  $S^3 \times S^1/Z_2$.

Furthermore, $PL^{(6)}$ can be identified with $M^{\#}$.
When $y^4 + y^5 \neq 0$, the corresponding element of $M^{\#}$  belongs to  $M^{(4)}$ (finite Minkowski space),
and is given by
\be
            x^\mu \,=\, \frac{y^\mu}{y^4 + y^5}\,,\quad  \mu \,=\,0,1,2,3.
\en

The ``light cone at infinity'' corresponds to the submanifold  $y^4 + y^5 = 0$ in $PL^{(6)}$.

\subsection{Conformal transformations in $Y^{(6)}$}
The $15$ conformal group generators {act via rotations} in the $(4+2)$-dimensional space $Y^{(6)}$ , so as to leave $PL^{(6)}$  {invariant}. Setting
\be
X_{mn} = y_m \partial_n - y_n \partial_m\,\quad (m < n)\,,
\en

\noindent
one has the $6$ rotation and boost generators
\be
M_{mn} = X_{mn} \quad (0 \leq m < n \leq 3)\,,
\en

the $4$ translation generators
\be
P_{m} = X_{m5} -  X_{m4}\,\quad (0 \leq m \leq 3)\,,
\en

the $1$ dilation generator
\be
D \,=\, - X_{45}\,,
\en

and the $4$ special conformal generators,
\be
K_m \,=\, - X_{m5} - X_{m4}\,,\quad  (0 \leq m < n \leq 3)\,.
\en

But of course, from these infinitesimal transformations we can only construct the \textit{special} conformal transformations, which act like (proper) rotations and boosts. Conformal inversion acts in $Y^{(6)}$ by {\it reflection} of the $y^5$ axis, which makes it easy to explore in other coordinate systems too:

\be \label{inversiony}
{y^\prime}^{\,m} = y^m (m = 0, 1, 2, 3, 4)\,, \quad {y^\prime}^5 = - y^5\,,
\en

\noindent
or more succinctly, ${y^\prime}^{\,m} = K^m_n y^n$, where $K^m_n = \mathrm{diag}\,[1,1,1,1,1,-1]$.

\subsection{The hexaspherical space $Q^{(6)}$}

This space is a different $(4+2)$-dimensional space, defined conveniently for dimensional reduction. For $q  \in \mathbf{R}^6$, write  $q = (q^a)$ with
    the index $a = 0,1,2,3,+,- $. Then define, for  $y \in Y^{(6)}$,
    with  $y^4 + y^5 \neq 0$,
\be
q^a  =  \frac{y^a}{y^4 + y^5}\, \,\,  (a = 0,1,2,3); \quad q^+ =  y^4 + y^5\,; \quad q^-  = \frac{y_m y^m}{(y^4 + y^5)^2}\,.
\en


In $Q^{(6)}$  the metric tensor is no longer flat:
\be
g_{ab}(q) = \begin{pmatrix}
              (q^+)^2 g_{\mu \nu} & 0 & 0 \\
              0 & q^- & \frac{q^+}{2} \\
              0 & \frac{q^+}{2}  & 0 \\
            \end{pmatrix}
\en

\noindent
The projective equivalence is simply

\be
  (q^0, q^1, q^2, q^3, q^+, q^-) \sim (q^0, q^1, q^2, q^3, \lambda q^+, q^-)\,,\quad \lambda \neq 0.
\en

We comment, however, that with this metric tensor, the map from contravariant to covariant vectors in $Q^{(6)}$ is actually two-to-one; hence it is not invertible. This suggests that one can improve on the hexaspherical coordinatization, a discussion we shall not pursue here.

When we take $q^-$ to zero, we have the light cone in $Q^{(6)}$, while fixing the value $q^+ = 1$ is one way to select a representative vector in each ray.
Another comment, however, is that fixing $q^+$ actually {\it breaks the conformal symmetry}. This is a subtle point that does not cause practical difficulty, but seems to have been unnoticed previously.

Convenient formulas for the transformation of $q$-coordinates under conformal inversion may be found in \cite{DupGolSht2014}.

\section[Maxwell theory with nonlinear constitutive equations]{Maxwell theory with nonlinear constitutive equations in $(4+2)$-dimensional spacetime}

\subsection{Nonlinear Maxwell equations in $Y^{(6)}$}

Next we introduce $6$-component fields $A_m$ in $Y^{(6)}$, and write
\be
F_{mn} \,=\, \partial_m A_n - \partial_n A_m\,,
\en

\noindent
so that
\be
\frac{\partial F_{mn}}{\partial y^k} + \frac{\partial F_{nk}}{\partial y^m} + \frac{\partial F_{km}}{\partial y^n} \,=\, 0\,.
\en

\noindent
While this is not really the {\it most\/} general possible ``electromagnetism'' in $4$ space and $2$ time dimensions, it is the theory most commonly discussed in the linear case, and the one we wish to generalize. As before, we defer writing the constitutive relations, and we have:
\be
\frac{\partial G^{mn}}{\partial y^m} \,=\, J^{\,n}\,,\label{constitutive}
\en

\noindent
where $J^{\,n}$ is the $6$-current.

For the nonlinear theory, we next need general conformal-invariant nonlinear constitutive equations relating $G^{mn}$ to $F_{mn}$. But {\it conformal invariance now means rotational invariance} in $Y^{(6)}$. Thus we write,
\be
G^{mn} \,=\, R^{mnk\ell}F_{n\ell} + P^{mnk\ell rs}F_{k\ell}F_{rs}\,,
\en

\noindent
where the tensors $R$ and $P$ take the general form,
\be
R^{mnk\ell} \,=\,r\,( \cdots)\,(\eta^{mk}\eta^{\ell n} - \eta^{nk}\eta^{\ell m})\,,\,\,\,\mathrm{and}
\,\,\,P^{mnk\ell rs} \,=\,p\,(\cdots)\,\epsilon^{mnk\ell rs}\,.
\en

\noindent
Here $r$ and $s$ must be functions of {\it rotational invariants}, which we next write down.

\subsection{Invariants for the general nonlinear Maxwell theory with conformal symmetry}

We can now write two rotation-invariant functionals of the field strength tensor in $Y^{(6)}$.
The first invariant is, as expected,
\be
I_1 \,=\, \frac{1}{2} F_{mn}F^{mn}\,.
\en

\noindent
But {unlike} in the $(3+1)$-dimensional case, the second rotational invariant is {\it trilinear} in the field strengths:
\be
I_2 \,=\,\frac{1}{2} \epsilon^{mnk\ell rs}F_{mn}F_{k\ell}F_{rs}\,.
\en

\noindent
This is a new pattern. Then, in Eq. (\ref{constitutive}), we have
\be
r \,=\, r\,(I_1,I_2)\,, \quad p \,=\, p\,(I_1,I_2)\,,
\en

\noindent
with $I_1$ and $I_2$ as above.

 In $Q^{(6)}$, the invariants take the form,
 \be
 I_1(q)= \frac{1}{2}\,F_{ab}(q)F^{ab}(q) = \frac{1}{2}\,g^{ac}g^{bd} F_{ab}(q)F_{cd}(q)\,,
 \en
 \be
 I_2(q) \,=\, \frac{1}{(q^+)^5}\,\epsilon^{abcdeg}F_{ab}(q)F_{cd}(q)F_{eg}(q)\,\quad
 \en
 \begin{equation}\nonumber
 \quad \quad =\, \frac{1}{2} (\mathrm{det}\,\bar{J})\,\epsilon^{abcdeg}F_{ab}(q)F_{cd}(q)F_{eg}(q)\,,
 \end{equation}

\noindent
 where $\bar{J}$ is a Jacobian matrix for transforming between $y$ and $q$-coordinates. Note that in the above, $\epsilon$ is the Levi-Civita {\it symbol}. The Levi-Civita {\it tensor} with raised indices is defined generally as $(1/\sqrt{|g|}\,)\epsilon$, where $g = \det [g_{ab}]$. Here this becomes $(\mathrm{det}\,\bar{J})\,\epsilon^{abcdeg}$.

The explicit presence of $q^+$ in the expression for $I_2$ explains why the condition $q^+ = 1$ does not respect the conformal symmetry: the value of $q^+$ can change under conformal transformations.

\section{Dimensional reduction to $(3+1)$ dimensions}

The final steps are to carry out a ray reduction and dimensional reduction of the $(4+2)$-dimensional Maxwell theory with conformal symmetry.

A {\it prolongation condition} states that the Maxwell fields respect the ray equivalence in $Y^{(6)}$):
\be
y^k \partial_k A_n \propto A_n\,.
\en

\noindent
A {\it splitting relation} allows the characterization of components tangential to $LC^{(6)}$:
\be
\frac{\partial A_n}{\partial y^n} \,=\,0\,\quad \mathrm{(a\,\,gauge\,\,condition)}\,.
\en

\noindent
One then expresses everything in $Q^{(6)}$ (hexaspherical coordinates), and restricts to the light cone by taking $q^- \to 0$, to obtain (as in the linear case) a general conformal nonlinear electromagnetism in $(3+1)$ dimensions, with some additional fields surviving the dimensional reduction.

In this article we have highlighted some new features suggested by the conformal symmetry of nonlinear Maxwell fields, including the idea of doubling the compactified Minkowski spacetime, and the trilinear form of one of the conformal invariant functionals in $Y^{(6)}$. For some additional details, see also \cite{DupGolSht2014}.

\section*{Acknowledgments}

The first author (GG) wishes especially to thank Professor Hounkonnou, who has been the inspiration behind his many visits to Benin as a visiting lecturer, and as a participant in the international ``Contemporary Problems in Mathematical Physics'' (Copromaph) conference series and school series. Acknowledgment is due to the International Centre for Theoretical Physics in Trieste, Italy, for partial support of these visits.

\end{document}